\documentclass[fleqn,usenatbib]{mnras}

\usepackage{newtxtext,newtxmath}

\usepackage[T1]{fontenc}

\DeclareRobustCommand{\VAN}[3]{#2}
\let\VANthebibliography\thebibliography
\def\thebibliography{\DeclareRobustCommand{\VAN}[3]{##3}\VANthebibliography}

\usepackage{graphicx}	
\usepackage{amsmath}	
\usepackage{footmisc}

\title[Non-thermal emission in SPT J2031]{Possible non-thermal origin of the hard X-ray emission in the merging galaxy cluster SPT-CL J2031--4037}

\author[M. S. Mirakhor et al.]{
M. S. Mirakhor$^{1\thanks{E-mail: msm0033@uah.edu}}$,
S. A. Walker$^{1}$, J. Runge$^{1}$, P. Diwanji$^{1}$
\\
$^{1}$Department of Physics and Astronomy, The University of Alabama in Huntsville, 301 Sparkman Drive, Huntsville, AL 35899, USA
}

\date{Accepted 2022 August 19. Received 2022 August 11; in original form 2022 March 27}
\pubyear{2015}

\begin{document}
\label{firstpage}
\pagerange{\pageref{firstpage}--\pageref{lastpage}}
\maketitle

\begin{abstract}
Non-thermal emission from clusters of galaxies at the high-energy X-ray regime has been searched with various instruments, but the detection significance of this emission has yet been found to be either marginal or controversial. Taking advantage of \textit{NuSTAR}'s unique capability to focus X-rays in the hard energy band, we present a detailed analysis of 238 ks \textit{NuSTAR} observations of the merging galaxy cluster SPT-CL J2031--4037, searching for non-thermal inverse Compton emission. Our spectral analysis of SPT-CL J2031--4037 shows a possibility that the hard X-ray emission of the cluster can be described by a non-thermal component, though we cannot completely rule out a purely thermal origin for this hard emission. Including the statistical and systematic uncertainties, our best model fit yields a 20--80 keV non-thermal flux of $3.93_{-1.10}^{+1.24} \times 10^{-12}$ erg s$^{-1}$ cm$^{-2}$. The estimated non-thermal flux is comparable to those found in other galaxy clusters using \textit{NuSTAR} and other X-ray instruments. Using this non-thermal flux with the existing radio data of the cluster, we estimate a volume-averaged magnetic field strength in the range of around 0.1--0.2 $\mu$G.


\end{abstract}




\section{Introduction}
\label{sec: intro}
It is well established that, mainly through radio observations, the intracluster medium (ICM) of galaxy clusters hosts a non-negligible fraction of relativistic particles and large-scale magnetic fields \citep[see][for a recent review]{vanWeeren2019}. In the presence of large-scale magnetic fields with a strength of a few $\mu$Gauss, the radio-emitting relativistic electrons emit synchrotron radiation, producing large-scale, diffuse structures known as radio halos and relics. The same population of relativistic electrons can also boost photon energies of the cosmic microwave background (CMB) to X-ray energies through inverse Compton (IC) scattering, resulting in a non-thermal high-energy tail in the X-ray spectrum of clusters \citep[e.g.][]{Rephaeli1979,Sarazin2000}.     

Measurements of non-thermal radiation provide valuable information that allows us to determine the mean magnetic field strength in the ICM directly, without invoking the assumption of energy equipartition between the field and relativistic electrons \citep{Rephaeli2008}. The magnetic fields play an important role in particle acceleration processes and in the energy distribution in the thermal gas through their effect on heat conduction, large-scale motions of the gas, and propagation of cosmic rays \citep[e.g.][]{Carilli2002,vanWeeren2019}. Cosmological simulations \citep[e.g.][]{ZuHone2011} predicted that the magnetic fields, in some cases, even play significant roles in the dynamics and structure of the ICM, such as in sloshing cool-core clusters, where the magnetic field is locally amplified in such a way that its pressure is near-equipartition with the thermal pressure of the ICM. Measuring non-thermal emission, therefore, would allow us to examine whether the average magnetic field strength in the ICM is large enough to impact the dynamical state of the thermal gas within the ICM.

Observationally, however, it is highly challenging to detect non-thermal IC emission, particularly below 10 keV, where thermal X-ray photons are very abundant \citep{Rephaeli2008}. At higher energies, where the thermal Bremsstrahlung emission falls off exponentially, non-thermal IC emission, in principle, can be observed as a hard X-ray excess on top of the thermal spectrum of clusters. 

The search for non-thermal emission started with the analysis of \textit{HEAO-1} data \citep{Rephaeli1987,Rephaeli1988}, but the first claim of non-thermal IC emission came after deep observations of the Coma cluster with the \textit{RXTE} and \textit{Beppo-SAX} satellites \citep{Rephaeli1999,Fusco-Femiano1999}. Several claims of non-thermal IC emission at energies above 20 keV have also been made in other galaxy clusters, although these detections were mostly found to be marginally significant and controversial \citep[see][for a review]{Rephaeli2008}.

\begin{table*}
\begin{minipage}{120mm}
    \centering
    \caption{\textit{NuSTAR} observations of SPT J2031}
    \begin{tabular}{lccccc}
    \hline
     Obs. ID   & RA     & Dec  & Obs. date   & \multicolumn{2}{c}{Unfiltered/Filtered exposure (ks)}\\
    \cline{5-6}
               &       &        &   & FPMA  &  FPMB  \\
        \hline
    70601001001 & 20 31 30.6 &	-40 39 32.0 & 2020-12-04 & 145.4/123.5 & 145.4/123.8 \\
    70601001003 & 20 32 01.4 &	-40 34 36.0 & 2021-05-27 & 92.8/75.9  & 92.8/71.7 \\
  \hline
    \end{tabular}
    \label{table: data}
\end{minipage}
\end{table*}

Observations with \textit{Suzaku} and \textit{Swift} satellites did not find evidence for the presence of non-thermal IC emission at levels that previously claimed with \textit{RXTE} and \textit{Beppo-SAX} \citep[e.g.][]{Ajello2009,Wik2012,Ota2014}. The only exception is the Bullet cluster \citep{Ajello2010}, although the non-thermal IC emission was found to be marginally significant. However, based on the \textit{NuSTAR}'s observations, the first focusing high-energy X-ray observatory that operates in the band between 3 and 79 keV, \citet{Wik2014} found that the global spectrum of the Bullet cluster at all energies can be explained by a multi-temperature thermal model, implying that the hard X-ray emission from the cluster has more likely a thermal origin.


Currently, the search for non-thermal IC emission in galaxy clusters with \textit{NuSTAR} is limited to a handful of clusters: the Bullet cluster \citep{Wik2014}, the Coma cluster \citep{Gastaldello2015}, Abell 523 \citep{Cova2019}, Abell 2163 \citep{Rojas2021}, and CL 0217$+$70 \citep{Tumer2022CL0217}. In this paper, taking advantage of \textit{NuSTAR}'s unique capability to focus X-rays in the hard energy band above 10 keV, we search for non-thermal emission in the merging galaxy cluster SPT-CL J2031--4037 (hereafter SPT J2031) using 238 ks \textit{NuSTAR} observations. \textit{NuSTAR} carries two identical co-aligned X-ray telescopes, labelled by their focal plane modules, FPMA and FPMB, with an effective area of $2 \times 350$ cm$^2$ at 9 keV and imaging half-power diameter of 58 arcsec. Its effective area is lower than that of previous X-ray instruments but is compensated by the focusing capability in the hard X-ray energy band, which greatly reduces point source contamination and the background level.

The galaxy cluster SPT J2031 is a massive ($M_{500}\sim 8\times 10^{14}$ M$_\odot$; \citealt{Chiu2018}), and morphologically disturbed \citep{Nurgaliev2017} galaxy cluster, situated at redshift $z \sim 0.3416$ \citep{Bohringer2004}. The cluster was first detected in the ROSAT-ESO Flux Limited X-ray (REFLEX) galaxy cluster survey \citep{Bohringer2004}, and later via the Sunyaev--Zel’dovich effect with the South Pole Telescope \citep[SPT;][]{Plagge2010,Williamson2011,Bleem2015}, and with the \textit{Planck} satellite \citep{Planck2016SZ}. Recently, \citet{Raja2020} reported the detection of diffuse radio emission in the galaxy cluster SPT J2031 with VLA (1.7 GHz) and GMRT (325 MHz) observations. The diffuse radio emission surrounds the central brightest cluster galaxy in SPT J2031, and the total extent of the radio emission at 325 MHz is around $0.8 \times 0.6$ Mpc, corresponding roughly to $2.7 \times 2.1$ arcmin at the redshift of the cluster. The spectral index of this diffuse radio emission at frequencies between 325 MHz and 1.7 GHz was found to be around 1.4. \citet{Raja2020} argued that the diffuse emission could be caused by a past merger event, which is in a transition phase into a mini halo. Given these interesting features, therefore, SPT J2031 is a promising target to search for non-thermal IC emission at high X-ray energies.


We describe the two \textit{NuSTAR} observations of SPT J2031 and the analyse procedure in Section \ref{sec: data}; the image and spectral analyses are presented in Sections \ref{sec: image} and \ref{sec: spectra}; our findings are discussed in Section \ref{sec: discussion}; and the summary of this work is presented in Section \ref{sec: summary}. Throughout this work, we adopt a $\Lambda$ cold dark matter cosmology with $\Omega_{\rm{m}}=0.3$, $\Omega_{\rm{\Lambda}}=0.7$, and $H_0=100\,h_{100}$ km s$^{-1}$ Mpc$^{-1}$ with $h_{100}=0.7$. We fixed the redshift of the cluster to the REFLEX value quoted above ($z = 0.3416$). Uncertainties are at the 90 per cent confidence level, unless otherwise stated. At the redshift of the cluster, 1 arcmin corresponds roughly to 294.5 kpc.

\section{Observations and data processing}
\label{sec: data}
The galaxy cluster SPT J2031 was observed by \textit{NuSTAR} on two occasions for a total unfiltered exposure of 238 ks (PI: S. A. Walker). The first observation was carried out on 2020 December 4 for an unfiltered exposure of 145.4 ks, and the second observation was carried out on 2021 May 27 for an unfiltered exposure of 92.8 ks. Table \ref{table: data} gives a summary of the \textit{NuSTAR} observations used in this work.

The data were reduced using standard pipeline processing (\textsc{HEASoft} v6.29 and \textsc{Nustardas} v2.1.1) and the 20210210 version of the \textit{NuSTAR} Calibration Database. We initiated the data processing by running the \textit{nupipeline} script to produce clean event files. The data were screened for periods of Earth occultations and enhanced background due to passages through the South Atlantic Anomaly (SAA). We adopted strict criteria regarding the SAA passages by setting the input parameter \textit{saamode} $=$ STRICT within the \textit{nupipeline} script. In addition, we set the input parameter '\textit{tentacle}' to 'yes' to flag time intervals in which the CZT detectors show an increase in event count rates when the observatory is entering into the SAA region. The filtered exposure time for both observations and telescopes is given in Table \ref{table: data}. 

\begin{figure}
\begin{center}
\includegraphics[width=1.\columnwidth]{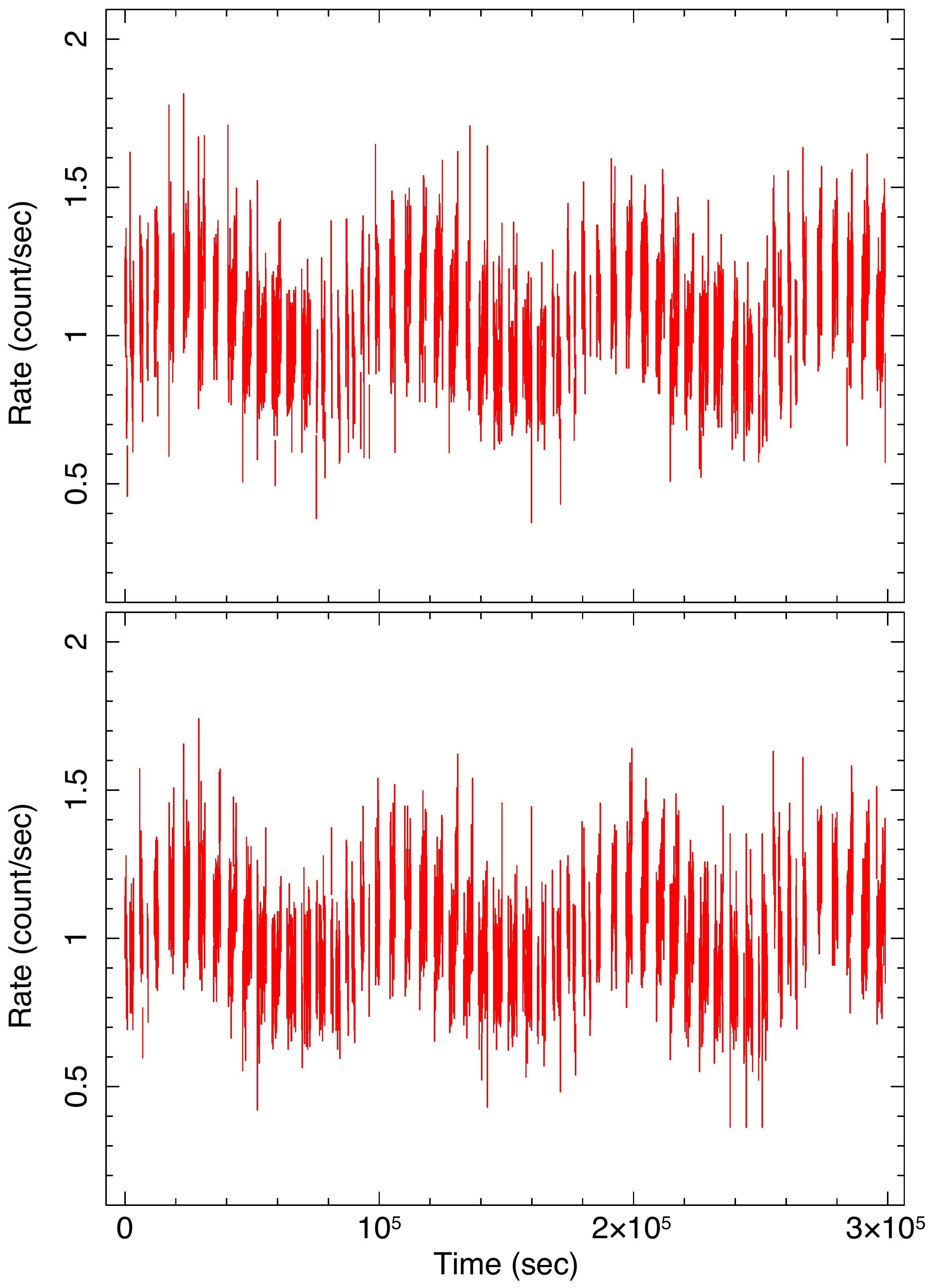}
\end{center}
\vspace{-0.5cm}
\caption{Light curves for the telescopes FPMA (\textit{top}) and FPMB (\textit{bottom}) of the observation 70601001001, extracted from the entire field of view using the full 1.6--160 keV energy range and the time binning of 100 s. There is no sign of notable fluctuations in the light curves. }
\label{fig: lc_01}
\vspace{-0.2cm}
\end{figure}

From the cleaned event files, we extracted images, light curves, and spectra by running the \textit{nuproducts} script. Exposure maps were generated using \textit{nuexpomap}, and the vignetting effects were taken into account. Within the call to \textit{nuproducts}, we set the input parameter '\textit{extended}' to 'yes' to generate response matrix and auxiliary response files appropriate for extended sources. We also visually inspected the cleaned event files of both observations and no notable fluctuations are found in the light curves. In Fig \ref{fig: lc_01}, as an example, we show the light curves of the telescopes FPMA and FPMB of the observation 7060100100.         


The \textit{NuSTAR} background, which varies to some extent both spectrally and spatially across the field of view, can generally be decomposed into four main components \citep{Wik2014}: the first component is an instrumental or internal background that dominates at energies above $\sim 20$ keV; the second component is an "aperture" background by stray light, which originates from unfocused cosmic X-rays that can leak through the aperture stops; the third component is reflected and scattered stray light, mainly from cosmic X-rays, the Earth, and the Sun, striking the detectors due to the open geometry of the observatory; and the fourth component is a focused cosmic background from the unresolved foreground and background sources within the field of view, which contributes significantly below 15 keV. In this work, we adopted the \textsc{idl} routines \textit{nuskybgd} suggested in \citet{Wik2014} to characterize the background and to generate scaled background spectra for the desired regions and images for the selected energy bands. For detailed description about the \textit{NuSTAR} background modelling see \citet{Wik2014}.

\begin{figure}
\begin{center}
\includegraphics[width=1.1\columnwidth]{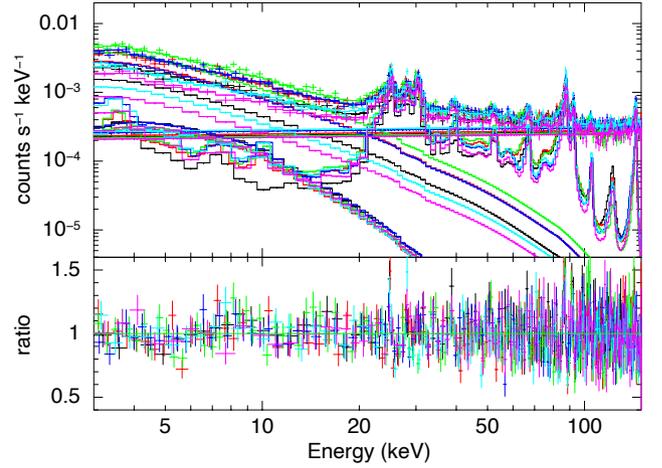}
\end{center}
\vspace{-0.5cm}
\caption{Upper panel: Fits to the background spectra of the observation 70601001001. The spectra of both telescopes FPMA and FPMB were extracted from the circular dashed regions shown in the left panel of Fig. \ref{fig: SPT_CLJ2031_img}. Different colors represent different background regions. Above 20 keV, the background is primarily due to cosmic ray-induced activation and instrumental fluorescence, along with instrumental particle background continuum. Between 5 and 20 keV, the aperture component is dominant. While the focused cosmic background component, which lies below the aperture lines, contributes mainly below 15 keV, the solar component, which is combined with the model of the instrumental lines, contributes significantly below 5 keV. Lower panel: The ratio of the data to the best-fitting model. A fit of equivalent quality is also obtained for the background of the observation 70601001003.  }  
\label{fig: bgd_01}
\vspace{-1. cm}
\end{figure}

\begin{figure}
\begin{center}
\includegraphics[width=1.1\columnwidth]{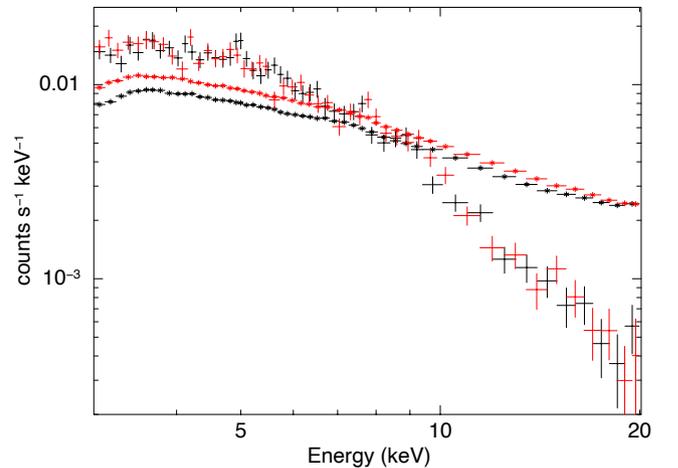}
\end{center}
\vspace{-0.5cm}
\caption{Background-subtracted global spectra of the observation 70601001001. The spectra from telescopes FPMA and FPMB are shown in black and red, respectively, and the two lower curves are their background spectra. The spectra are background dominated above 10 keV. } 
\label{fig: src_bgd_01}
\vspace{-0.2cm}
\end{figure}

For both observations, we obtained fits of equivalent quality to the background spectra. As an example, in Fig. \ref{fig: bgd_01}, we show results of our fits to the background spectra of the observation 70601001001. Fig. \ref{fig: src_bgd_01} shows the background-subtracted global spectra of the observation 70601001001. The spectra from telescopes FPMA and FPMB are shown in black and red, respectively, and the two lower curves are their background spectra. The spectra of both telescopes were extracted from the circular dashed regions shown in the left panel of Fig. \ref{fig: SPT_CLJ2031_img}, and they were fitted independently, except for the focused X-ray cosmic background, where their normalizations were tied together.


As noted from Fig. \ref{fig: bgd_01}, at energies above 20 keV, the background is primarily due to instrumental fluorescence and cosmic ray-induced activation from the SAA passages, along with instrumental particle background continuum. From 5--20 keV, the aperture component dominates the background, while the solar component, which is combined with the model of the instrumental lines, contributes significantly below 5 keV. The focused cosmic background component, which lies below the aperture lines, contributes mainly below 15 keV.

\section{Image analysis}
\label{sec: image}
The analysis procedures described in Section \ref{sec: data} were applied on both observations and co-aligned X-ray telescopes, FPMA and FPMB. We used \textsc{HEASoft}'s \textit{ximage} package to combine the count, background, and exposure images from both observations and telescopes. We carried out exposure correction by creating exposure maps at single energies for each band, which roughly corresponds to the mean emission-weighted energy of the band \citep[e.g.][]{Gastaldello2015,Rojas2021}.

\begin{figure*}
\begin{center}
\includegraphics[width=\textwidth]{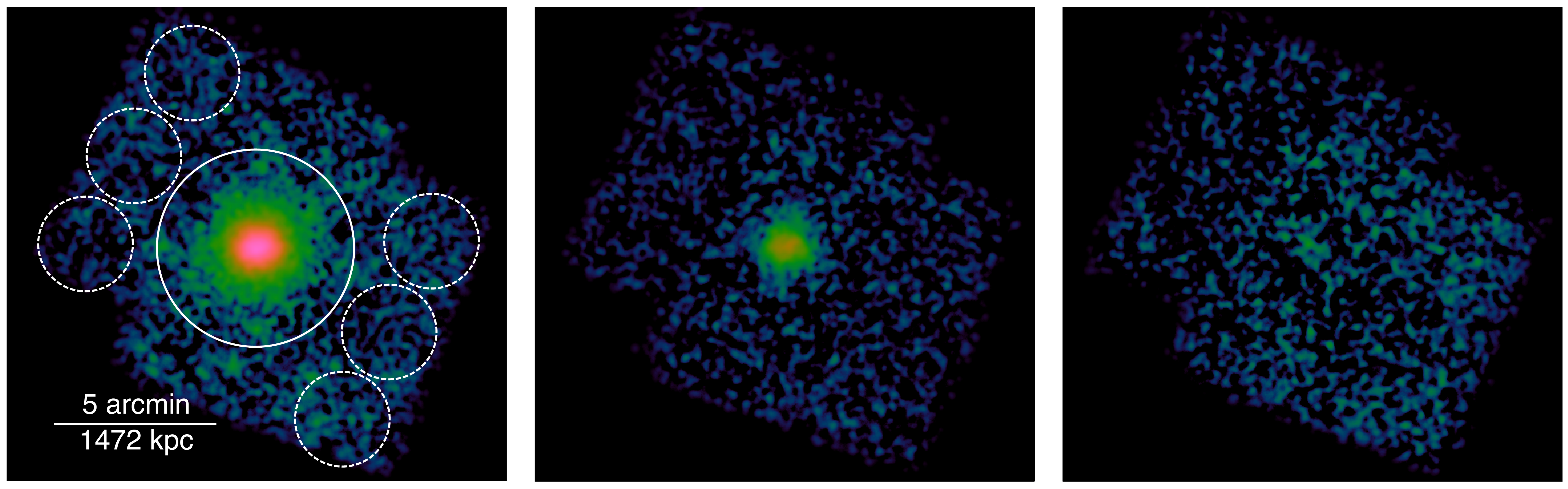} 
\end{center}
\caption{Background-subtracted and exposure-corrected mosaicked images of the galaxy cluster SPT J2031, combining both observations and telescopes. The images were smoothed by a Gaussian kernel with $\sigma=17.2$ arcsec (7 pixels) to be consistent with the \textit{NuSTAR}'s PSF of $\sim 18$ arcsec FWHM. The energy band for each image is from left to right: 3--10 keV, 10--20 keV, and 20--50 keV. The white circle (5 arcmin radius) marks the region in which the source spectra were extracted. The dashed circles mark the regions from which the background spectra were extracted.}
\label{fig: SPT_CLJ2031_img}
\vspace{-0.2cm}
\end{figure*}

Fig. \ref{fig: SPT_CLJ2031_img} shows the background-subtracted and exposure-corrected mosaicked images of the cluster SPT J2031 in in three energy bands: 3--10 keV (left-hand panel), 10--20 keV (middle panel), and 20--50 keV (right-hand panel). The images are smoothed by a Gaussian kernel with $\sigma=17.2$ arcsec (7 pixels) to be consistent with the \textit{NuSTAR}'s point spread function (PSF) of $\sim 18$ arcsec FWHM (full width at half-maximum). While the 3--10 keV and 10--20 keV images show the distribution of the hot gas in the field of SPT J2031, there is no evidence for the cluster emission in the 20--50 keV image, which appears to be completely background dominated.

We inspected the mosaicked images of SPT J2031 for point source contamination, and we detected few bright point sources in the 3--10 keV image. These sources were then excluded from further analysis. Above 10 keV, we did not detect any bright point sources in the SPT J2031 field.


\section{Spectral analysis}
\label{sec: spectra}
Based on Fig. \ref{fig: src_bgd_01}, the spectra are background dominated above 10 keV. To include the effect of systematic uncertainties due to the background fluctuations, we follow the procedure described in \citet{Wik2014}. For each spectrum, we simulated 1000 different realizations of the background, adopting the systematic errors derived in \citet{Wik2014} for the various background components (3 per cent for the instrumental component, 8 per cent for the aperture component, 42 per cent for the focused cosmic component, and 10 per cent for the Solar component). Assuming that the fluctuations follow a Gaussian distribution, each realization of the generated backgrounds is subtracted from the source spectrum, and systematic uncertainties associated with various spectral models are then estimated.



\begin{figure}
\begin{center}
\includegraphics[width=1.1\columnwidth]{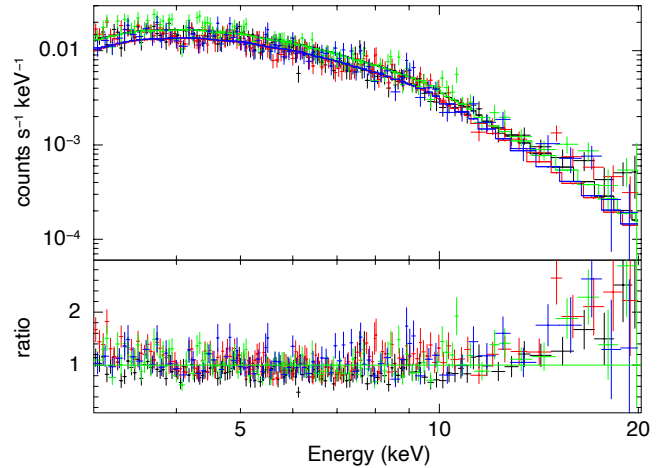} 
\end{center}
\caption{Background-subtracted 3--20 keV spectra of the galaxy cluster SPT J2031 extracted from  a circular region of radius 5 arcmin shown in the left-hand panel of Fig. \ref{fig: SPT_CLJ2031_img}. The spectra were fitted to a single-temperature model. A significant excess is visible between 10 and 20 keV. } 
\label{fig: 1T_3to20}
\vspace{-0.2cm}
\end{figure}

\subsection{Global spectrum}
\label{sec: global}
\begin{figure*}
\begin{center}
\includegraphics[width=\textwidth]{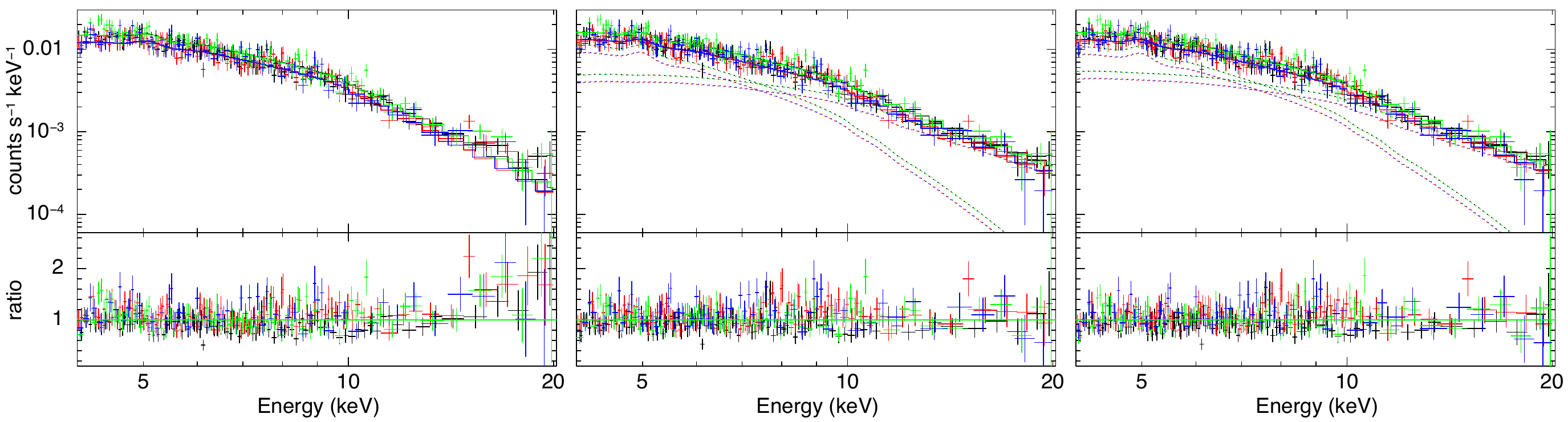} 
\end{center}
\caption{\textit{Upper panels}: 4--20 keV global spectra of both observations and telescopes of SPT J2031, extracted from a circular region of radius 5 arcmin shown in the left-hand panel of Fig. \ref{fig: SPT_CLJ2031_img}. The spectra were fitted to a single-temperature model (\textit{left}), two-temperature model (\textit{middle}), and single-temperature plus power-law model with a free photon index (\textit{right}). The solid lines are the best-fitting models to the spectra. The components of the two-temperature and single-temperature plus power-law models are shown as dashed lines, with the IC component (\textit{right}) dominating at energies above $\sim 10$ keV. \textit{Lower panels}: The ratio of the data to the best-fitting models. The two-temperature model provides a good fit to the spectral data,  but a fit of equivalent quality is also obtained with the model that includes a power-law component. } 
\label{fig: global_spectrum}
\vspace{-0.2cm}
\end{figure*}

The global spectrum of the SPT J2031 cluster was extracted from a circular region of radius 5 arcmin, centred on the location of the peak of the cluster's surface brightness (Fig. \ref{fig: SPT_CLJ2031_img}, left-hand panel). The spectral data of both observations and telescopes were simultaneously fitted with a single-temperature (1T) model, consisting of an APEC thermal component \citep{Smith2001} with Galactic absorption fixed at $N_{\rm{H}}=3.21 \times 10^{20}$ cm$^{-2}$ and the redshift fixed at $z = 0.3416$. The spectral fitting was first carried out in the 3--20 keV energy band using the standard X-ray spectral fitting package \citep[\textsc{xspec};][]{Arnaud1996}. The values of best-fitting parameters are obtained using C-statistic (C-stat), which implements the W-statistic, a modification of the Cash statistic. The best fit yields a temperature $T = 9.2_{-0.3}^{+0.4}$ keV and a metal abundance $Z = 0.03_{-0.02}^{+0.02}$ Z$_\odot$ at 90 per cent confidence levels. The C-stat value associated with the 1T model fit is 1903.1 with 1693 degrees of freedom (dof). The best-fitting parameters are shown in Table \ref{table: model_comparison}, and the best-fitting model to the global spectra extracted from both observations and telescopes is shown in Fig. \ref{fig: 1T_3to20}. As shown in this figure, a significant excess is visible between 10 and 20 keV.



The 1T model that we considered above provides the simplest possible description of the spectral data, and it is unlikely to be a realistic description, particularly for morphologically disturbed clusters like SPT J2031 \citep[see][]{McDonald2013,Nurgaliev2017}. To provide a more realistic description of the data, we fitted the spectra in the 3--20 keV energy band with a two-temperature (2T) model, consisting of two APEC thermal components with their metal abundances tied together. The best fit of the 2T model to the X-ray spectra gives temperatures $T_1 = 2.4_{-0.5}^{+1.7}$ keV and $T_2 = 18.3_{-2.2}^{+13.7}$ keV, and an abundance $Z = 0.10_{-0.05}^{+0.05}$ Z$_\odot$ at 90 per cent confidence levels (see Table \ref{table: model_comparison}). The C-stat value associated with this fit is 1774.8 with 1691 dof, indicating a better fit than the 1T model fit. However, the best-fitting temperatures of the 2T model are different from each other by an order of magnitude and bracketing the ends of the 3--20 keV bandpass. In addition, both temperature measurements are poorly constrained.




Another approach is to fit the X-ray spectra in the 3--20 keV band to a single-temperature plus power-law (1T $+$ IC) model. Based on the spectral index of the SPT J2031 radio halo that was found to be around 1.4 between frequencies 325 MHz and 1.7 GHz \citep{Raja2020}, the photon spectral index of the power-law component was fixed at $\Gamma = 2.4$. The values of the best-fitting temperature and abundance yielded from this fit are $T = 11.5_{-4.4}^{+7.1}$ keV and $Z = 0.48_{-0.25}^{+0.31}$ Z$_\odot$ at 90 per cent confidence levels (Table \ref{table: model_comparison}). The best fit of the 1T $+$ IC model yielded a 20--80 keV flux of $1.68_{-0.18}^{+0.22} \times 10^{-12}$ erg s$^{-1}$ cm$^{-2}$ on the non-thermal component. With this steep power-law index, we find that the emission measure of the thermal component drops by a factor of around 20 from the 2T model to the 1T $+$ IC model, and the abundance increases by a factor of around 4. Furthermore, the emission measure of the thermal component in the 1T $+$ IC model is lower by a factor of around 6 than the emission measure of the IC component. This essentially implies that all of the emission above 10 keV and most below 10 keV has a non-thermal origin. Generally, one would expect the thermal gas to be dominant in the 3--10 keV band, where most of the photons reside. Statistically, the 2T model provides a better fit than the 1T $+$ IC model when the power-law photon index is fixed at 2.4, as inferred by their C-stat/dof values (1774.8/1691 for the 2T model compared to 1780.4/1692 for the 1T $+$ IC model).




We also repeated the fitting process by fitting the 1T $+$ IC model to the global spectrum in the 3--20 keV band, but this time we allowed the photon spectral index of the power-law component to be free. The best-fitting model yielded a temperature  $T = 2.1_{-0.7}^{+2.6}$ keV and abundance $Z = 0.44_{-0.22}^{+0.24}$ Z$_\odot$ at 90 per cent confidence levels. The best fit yielded an index of $2.3_{-0.2}^{+0.1}$ and a 20--80 keV flux of $2.31_{-0.22}^{+0.26} \times 10^{-12}$ erg s$^{-1}$ cm$^{-2}$ on the non-thermal component.  The comparison between the 1T $+$ IC model with a free photon index and the 2T model points to a slightly better fit with the former model (1769.3/1691 for the 1T $+$ IC model compared to 1774.8/1691 for the 2T model). However, the 1T $+$ IC model with a free photon index has an unphysically low thermal component at 2.1 keV. As shown in Table \ref{table: model_comparison}, the emission measure of the thermal component drops by a factor of 6 from the 2T model to the 1T $+$ IC model, and the abundance increases by almost a factor of around 4. In addition, the emission measure of the thermal component in the 1T $+$ IC model is lower by a factor of around 1.5 than that of the IC component, implying that the IC component is dominating the entire spectrum, i.e. the X-ray emission at all energies is dominated by non-thermal emission. This, in turn, makes this model physically unrealistic.

Recently, \citet{Madsen2020} have found that the effective area of the FPMA detector could be slightly off below $\sim 5$ keV due to thermal blanket tear in the \textit{NuSTAR} optic aligned with that detector, allowing additional low energy photons through. In Fig. \ref{fig: 1T_3to20}, there is a slight excess in the 1T residuals below 4 keV that the low-temperature component may be fitting to. If that's the case, then the parameter values are likely to be quite different in the 2T and T $+$ IC models if the fits are carried out with data from 4--20 keV.

Therefore, we repeated the fits but only  with the global X-ray data from 4--20 keV. The values of the best-fitting parameters obtained from the best 1T, 2T, and 1T $+$ IC model fits to the X-ray spectra are shown in Table \ref{table: model_comparison_4to20}. The best 1T model fit to the X-ray spectra is shown in the left-hand panel of Fig. \ref{fig: global_spectrum}. Overall, the 1T model provides a good fit to the X-ray spectra below 10 keV, but there is a significant excess between 10 and 20 keV. The abundance is now more consistent with the abundances measured in smaller regions (and in clusters generally), suggesting a more accurate fit was obtained and that the lower energy data below 4 keV were likely biasing the measurement. Our results show that the statistics improve by adding a thermal or power-law component to the 1T model. When it comes to the comparison of the 1T $+$ IC and 2T models, our findings suggest that the 2T model provides a good fit to the data over the entire spectrum, but a fit of equivalent quality is also obtained with the 1T $+$ IC model with a free photon index (C-stat/dof $=$ 1618.5/1591 for the 2T model compared to C-stat/dof $=$ 1618.4/1591 for the 1T $+$ IC model).

As shown in Table \ref{table: model_comparison_4to20}, the temperature obtained from the 1T $+$ IC model with a free photon index is $5.9_{-1.0}^{+2.1}$ keV, which is consistent with the lower-temperature component of the 2T model, and is around $1\sigma$ lower than the temperature obtained from the 1T model. In the middle and right-hand panels of Fig. \ref{fig: global_spectrum}, we show the best-fitting 2T and 1T $+$ IC models to the X-ray spectra in the 4--20 keV band. This figure shows that the non-thermal component of the 1T $+$ IC model with a free photon index mimics the higher-temperature component of the 2T model, while its thermal component mimics the lower-temperature component of the 2T model. Overall, our results suggest that, if the photon spectral index is allowed to be free, the 1T $+$ IC and 2T models fit the spectral data equally well.    




\begin{table*}
\begin{minipage}{175mm}
    \centering
    \caption {Results of the global spectral analysis in the 3--20 keV energy band. The statistical uncertainties are at the 90 per cent confidence levels, followed by the 90 per cent systematic uncertainties.}
    \begin{tabular}{lcccccccc}
   \hline
    Model  & $T_1$  & $Z$ & Norm$_1$\footnote{\label{note1}Normalization of the APEC thermal component. In models with an IC component, Norm$_2$ is the normalization of the power-law component, which is given in units of photons keV$^{-1}$ cm$^{-2}$ s$^{-1}$ at 1 keV (10$^{-3}$).} & $T_2$ & $\Gamma$ & Norm$_2$\footref{note1} & IC flux\footnote{\label{note2}Non-thermal flux in the 20--80 keV band.} & C-stat/dof \footnote{\label{note3}Associated uncertainties are distribution of C-stat values using 1000 realizations of the background.}   \\
          & (keV) &  (Z$_\odot$)  & ($10^{-3}$ cm$^{-5}$)  & (keV)  &  & ($10^{-3}$ cm$^{-5}$) & ($10^{-12}$ erg s$^{-1}$ cm$^{-2}$) &   \\    \hline
1T & $9.2_{-0.3-0.3}^{+0.4+0.3}$ & $0.03_{-0.02-0.01}^{+0.02+0.01}$ & $7.47_{-0.25-0.14}^{+0.25+0.13}$ & --- & --- & --- & --- & $1903.1_{-182}^{+202}/1693$ \\ 

2T & $2.4_{-0.5-1.0}^{+1.7+1.1}$ & $0.10_{-0.05-0.02}^{+0.05+0.01}$  & $11.59_{-0.78-0.40}^{+0.81+0.28}$  & $18.3_{-2.2-2.5}^{+13.7+9.1}$ & --- & $4.03_{-0.66-0.35}^{+0.45+0.30}$ & --- & $1774.8_{-172}^{+199}/1691$ \\ 

1T $+$ IC  & $11.5_{-4.4-2.6}^{+7.1+3.6}$ & $0.48_{-0.25-0.02}^{+0.31+0.02}$ & $0.56_{-0.09-0.08}^{+0.10+0.08}$ & --- & 2.4 (fixed) & $3.14_{-0.32-0.20}^{+0.23+0.20}$ & $1.68_{-0.18-0.28}^{+0.22+0.30}$ & $1780.4_{-175}^{+198}/1692$  \\

1T $+$ IC  & $2.1_{-0.7-0.9}^{+2.6+1.0}$ & $0.44_{-0.22-0.03}^{+0.24-0.03}$ & $1.88_{-0.31-0.18}^{+0.37+0.21}$ & --- & $2.3_{-0.2-0.1}^{+0.1+0.1}$ (free) & $2.66_{-0.36-0.20}^{+0.35+0.19}$ & $2.31_{-0.22-0.43}^{+0.26+0.48}$ & $1769.3_{-170}^{+191}/1691$  \\

8T  & --- & --- &$0.92 \pm0.02$\footnote{\label{note4}Overall normalization constant for the 8 thermal components.} & --- & --- & --- & --- & $1900.1_{-174}^{+197}/1695$ \\

8T $+$ IC   & --- & --- & --- & --- & 2.4 (fixed) & $1.95_{-0.13-0.10}^{+0.17+0.08}$ & $1.52_{-0.22-0.32}^{+0.23+0.31}$ & $1787.8_{-174}^{+199}/1694$ \\

8T $+$ IC   & --- & --- & --- & --- & $2.3_{-0.1-0.1}^{+0.1+0.1}$ (free) & $2.88_{-0.27-0.19}^{+0.21+0.17}$ & $1.79_{-0.20-0.34}^{+0.20+0.37}$ & $1783.8_{-173}^{+196}/1693$ \\

  \hline
    \end{tabular}
    \vspace{-5mm}
    \label{table: model_comparison}
\end{minipage}
\end{table*}

\begin{table*}
\begin{minipage}{175mm}
    \centering
    \caption {Same as Table \ref{table: model_comparison}, except the global spectral fitting is carried out in the 4--20 keV energy band.}  
    \begin{tabular}{lcccccccc}
   \hline
    Model  & $T_1$  & $Z$ & Norm$_1$ & $T_2$ & $\Gamma$ & Norm$_2$ &IC flux & C-stat/dof   \\
          & (keV) &  (Z$_\odot$)  & ($10^{-3}$ cm$^{-5}$)  & (keV)  &  & ($10^{-3}$ cm$^{-5}$) & ($10^{-12}$ erg s$^{-1}$ cm$^{-2}$) &   \\    \hline
1T & $10.6_{-0.6-0.6}^{+0.6+0.7}$ & $0.17_{-0.07-0.04}^{+0.08+0.02}$ & $6.03_{-0.29-0.20}^{+0.32+0.19}$ & --- & --- & --- & --- & $1670.9_{-166}^{+186}/1593$ \\ 

2T & $5.3_{-0.9-1.1}^{+1.6+1.4}$ & $0.11_{-0.04-0.01}^{+0.05+0.01}$  & $6.76_{-0.76-0.88}^{+0.77+0.86}$  & $18.1_{-5.3-1.4}^{+14.4+10.8}$ & --- & $1.95_{-0.29-0.30}^{+0.62+0.88}$ & --- & $1618.5_{-164}^{+177}/1591$ \\ 

1T $+$ IC  & $12.6_{-1.4-3.7}^{+8.8+5.2}$ & $0.46_{-0.21-0.04}^{+0.11+0.03}$ & $1.17_{-0.41-0.65}^{+0.44+0.64}$ & --- & 2.4 (fixed) & $2.68_{-0.67-0.87}^{+0.38+0.77}$ & $1.36_{-0.15-0.24}^{+0.16+0.27}$ & $1633.5_{-171}^{+183}/1592$  \\

1T $+$ IC  & $5.9_{-1.0-0.6}^{+2.1+0.3}$ & $0.11_{-0.05-0.02}^{+0.04-0.02}$ & $6.39_{-0.11-0.82}^{+0.11+0.67}$ & --- & $1.6_{-0.6-0.3}^{+0.7+0.3}$ (free) & $0.26_{-0.03-0.09}^{+0.05+0.09}$ & $3.93_{-0.65-0.45}^{+0.76+0.48}$ & $1618.4_{-165}^{+179}/1591$  \\

8T  & --- & --- & $0.92 \pm0.02$  & --- & --- & --- & --- & $1665.8_{-168}^{+184}/1595$ \\

8T $+$ IC   & --- & --- & --- & --- & 2.4 (fixed) & $7.20_{-1.21-0.69}^{+1.15+0.44}$ & $3.53_{-0.17-0.28}^{+0.19+0.30}$ & $1638.4_{-173}^{+184}/1594$ \\

8T $+$ IC   & --- & --- & --- & --- & $2.3_{-0.1-0.1}^{+0.1+0.1}$ (free) & $5.30_{-1.01-0.26}^{+0.94+0.38}$ & $3.74_{-0.19-0.32}^{+0.18+0.40}$ & $1634.5_{-170}^{+181}/1593$ \\

  \hline
    \end{tabular}
    \label{table: model_comparison_4to20}
\end{minipage}
\end{table*}

\subsection{Thermodynamic maps}
\label{sec: maps}
\begin{figure}
\begin{center}
\includegraphics[width=1.\columnwidth]{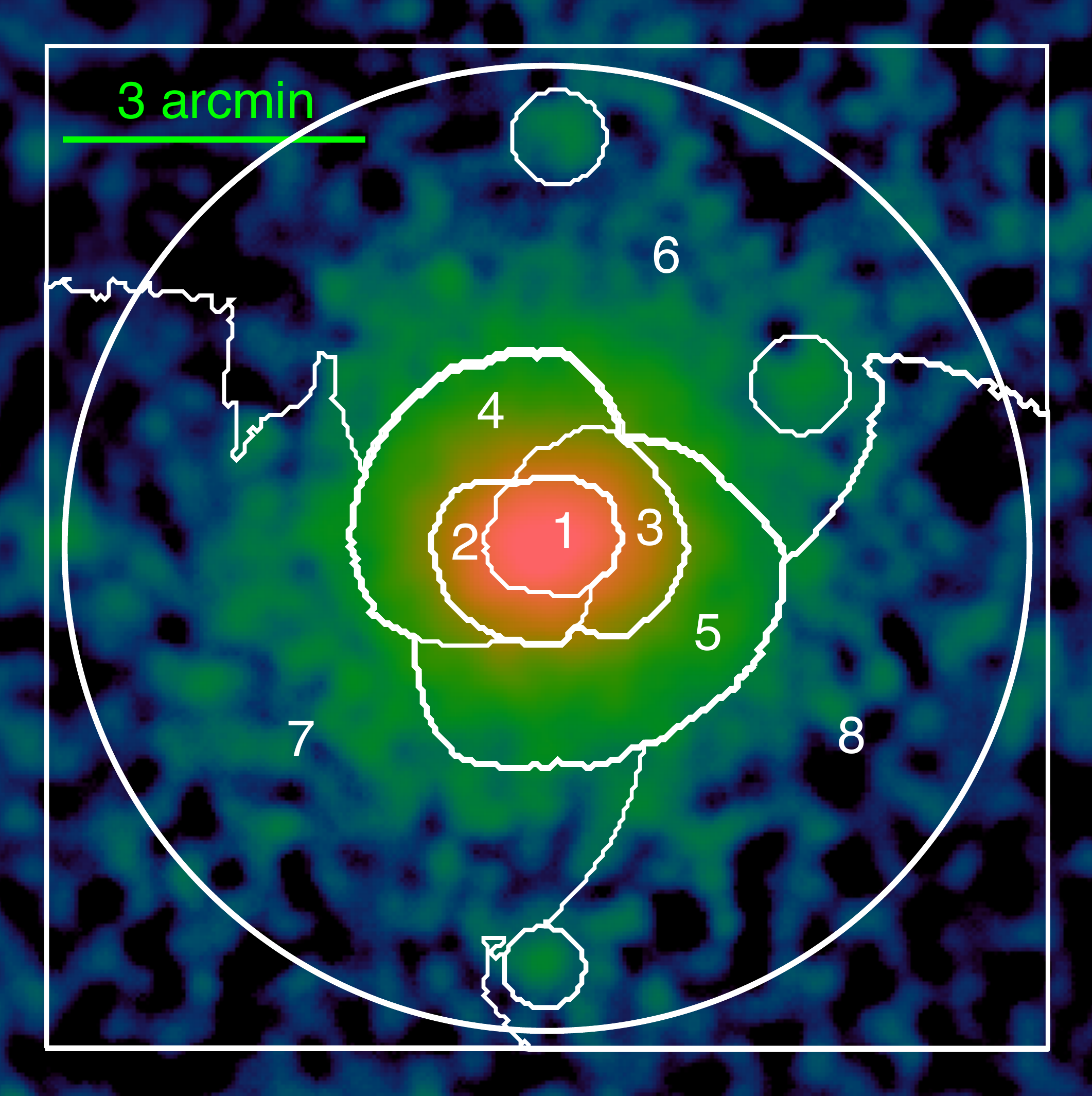}
\end{center}
\caption{Regions obtained using the contour binning technique, overlaid on the \textit{NuSTAR} image of the SPT J2031 cluster in the 3--10 keV energy band. For each selected region, the X-ray spectrum was extracted and fitted to a single-temperature model, using the data from both observations and telescopes.}
\label{fig: bins}
\vspace{-0.2cm}
\end{figure}

\begin{figure*}
\begin{center}
\includegraphics[width=0.535\textwidth]{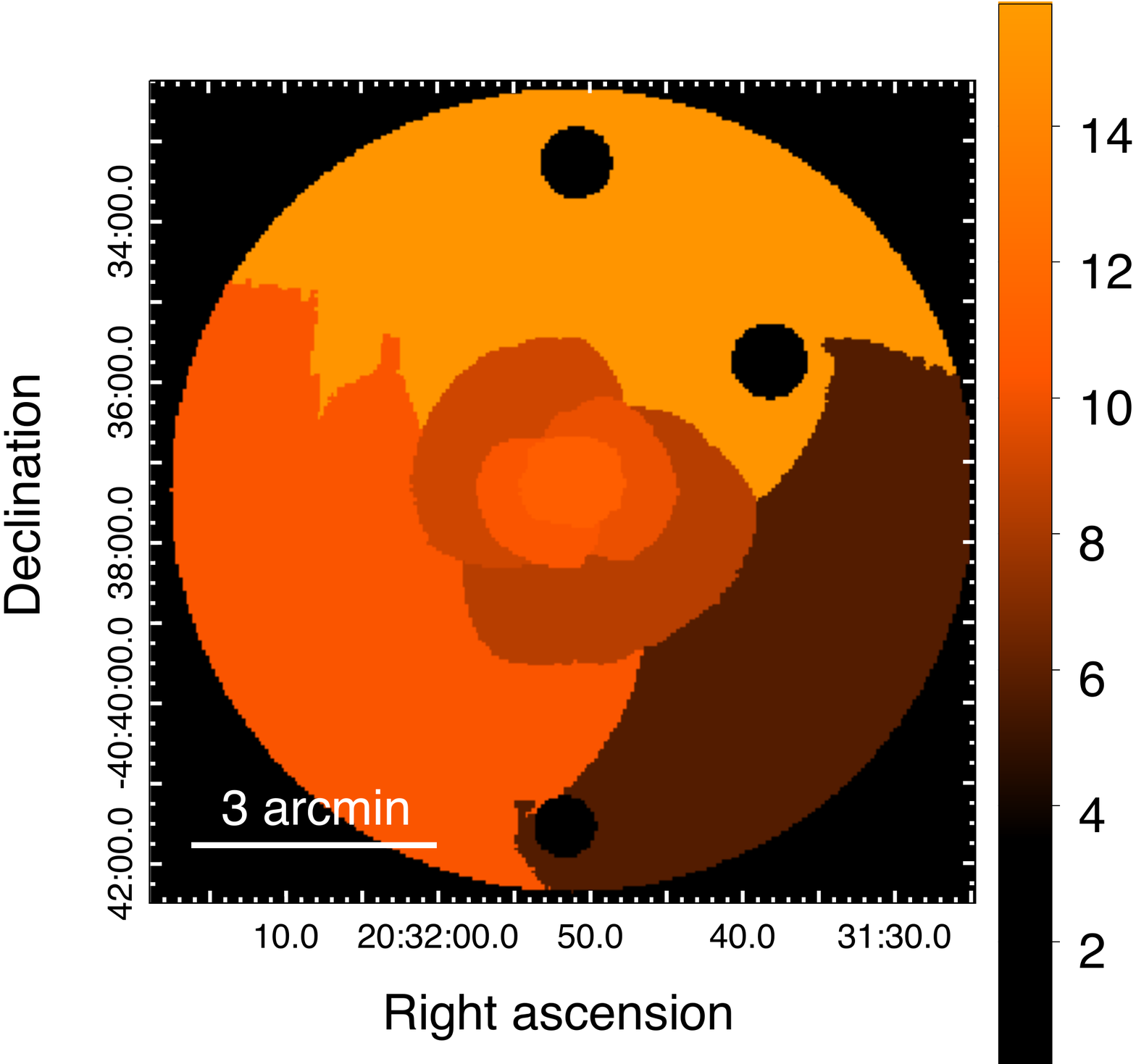}
\hspace{-1.4cm}
\includegraphics[width=0.535\textwidth]{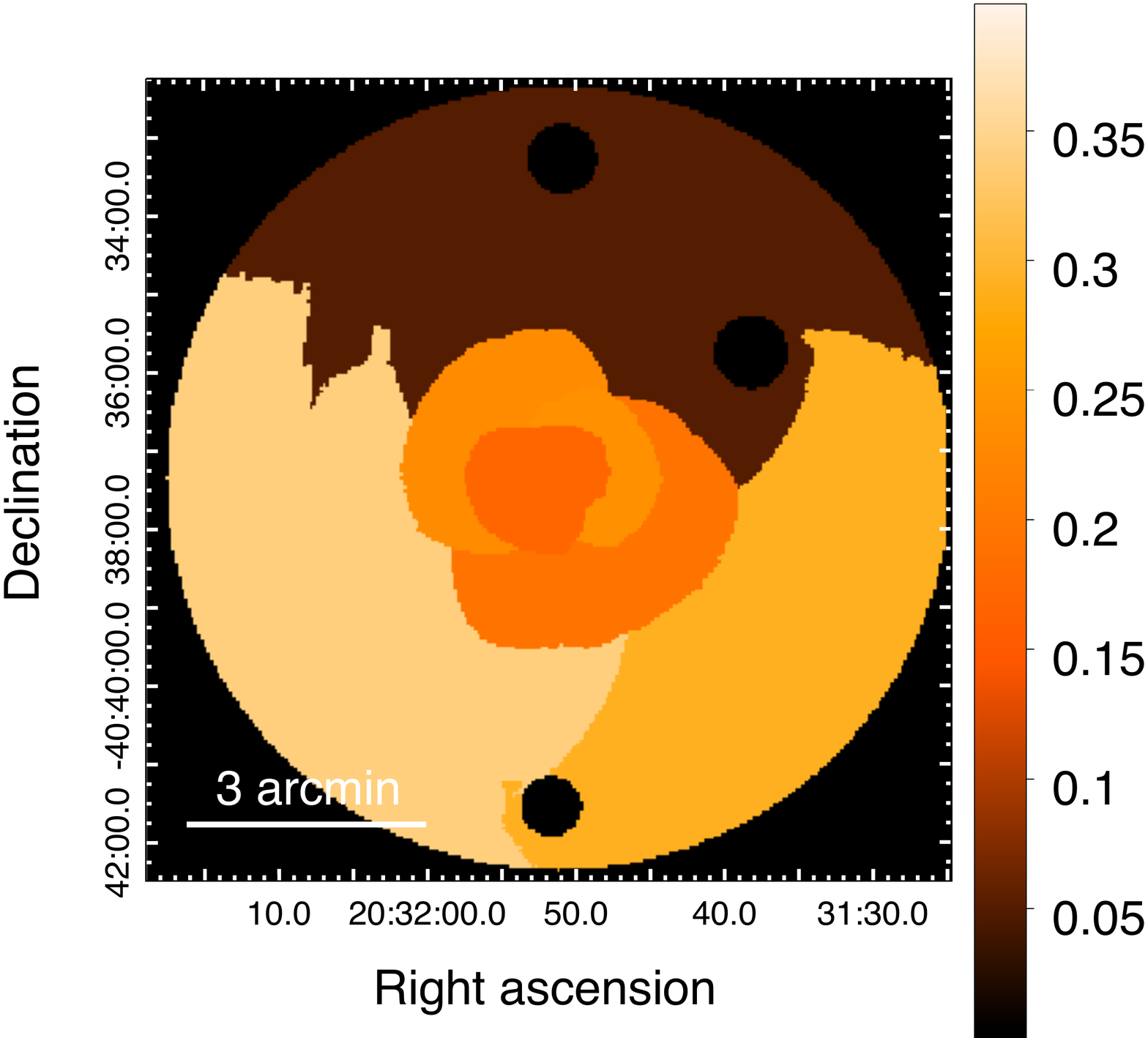}
\end{center}
\caption{Temperature (\textit{left}) and iron abundance (\textit{right}) maps of the SPT J2031 cluster, consisting of 8 regions obtained using the contour binning technique. The colour scales are in units of keV (\textit{left}) and Z$_\odot$ (\textit{right}).}
\label{fig: temp_abund_maps}
\vspace{-0.2cm}
\end{figure*}

A proper description of the properties of dynamically disturbed galaxy clusters like SPT J2031 requires a more detailed spatial analysis. To do that, we adopted the contour binning software \textsc{contbin} version 1.6 \citep{Sanders2006contour} to choose spatial regions for X-ray spectroscopy, using a threshold for the signal-to-noise ratio of 50. The software chooses regions using contours from adaptively smoothed images in such a way that the generated regions closely follow the surface brightness. Given \textit{NuSTAR}'s large PSF of $\sim 18$ arcsec FWHM, we selected 8 regions, covering the field of the SPT J2031 cluster.

In Fig. \ref{fig: bins}, we show regions obtained using the contour binning technique, overlaid on the \textit{NuSTAR} image of the SPT J2031 cluster in the 3--10 keV energy band. For each selected region, the X-ray data from the 3--20 keV and 4--20 energy bands were fitted to a 1T model using \textsc{xspec}, following the procedure described in Section \ref{sec: global}. We find that the best-fitting parameters from both energy bands are consistent well with each other. In Table \ref{table: single_temp_model}, we show the values of the best-fitting parameters to the X-ray data in the 4--20 keV band for each selected region. As inferred by their C-stat/dof values, the 1T model, overall, provides a reasonable fit to the spectra extracted from the selected regions. The resulting temperature and metal abundance maps of SPT J2031 are shown in Fig. \ref{fig: temp_abund_maps}.


\begin{table}
\begin{minipage}{80mm}
    \centering
    \caption{1T model fit to the 4--20 keV spectral data for the 8 selected regions.}
    \begin{tabular}{lcccc}
   \hline
    Region & $T$  & $Z$ & Norm\footnote{\label{note5}Normalization of the APEC thermal component.}  & C-stat/dof   \\
          & (keV) &  (Z$_\odot$)  & ($10^{-3}$ cm$^{-5}$)  &   \\    \hline
1 & $10.9_{-0.8}^{+1.0}$ & $0.17_{-0.08}^{+0.10}$ & $1.16_{-0.10}^{+0.11}$ & 1613.6/1593\\
2 & $10.2_{-1.1}^{+1.3}$ & $0.17_{-0.09}^{+0.11}$ & $0.69_{-0.09}^{+0.09}$ & 1675.1/1593\\
3 & $9.8_{-1.0}^{+1.3}$ & $0.24_{-0.12}^{+0.14}$ & $0.60_{-0.08}^{+0.04}$ & 1712.4/1593\\
4 & $9.0_{-1.2}^{+1.6}$ & $0.23_{-0.12}^{+0.14}$ & $0.61_{-0.08}^{+0.09}$ & 1800.3/1593\\
5 & $8.4_{-1.1}^{+1.4}$ & $0.19_{-0.11}^{+0.13}$ & $0.63_{-0.09}^{+0.10}$ & 1763.3/1593\\
6 & $15.7_{-7.1}^{+14.8}$ & $0.05_{-0.04}^{+0.06}$ & $0.47_{-0.71}^{+0.38}$ & 1920.4/1593\\
7 & $10.2_{-2.7}^{+4.3}$ & $0.34_{-0.18}^{+0.20}$ & $0.77_{-0.11}^{+0.12}$ & 1901.3/1593\\
8 & $5.7_{-1.6}^{+2.3}$ & $0.29_{-0.13}^{+0.15}$ & $0.35_{-0.11}^{+0.11}$ & 1761.6/1593\\
  \hline
    \end{tabular}
    \vspace{-5mm}
    \label{table: single_temp_model}
\end{minipage}
\end{table}

Following the previous work \citep[e.g.][]{Gastaldello2015,Cova2019,Rojas2021}, we summed the 8 1T (8T) models with temperature, abundance, and normalization fixed at their best-fitting values. We only allowed for the overall normalization parameter to be free to account for possible discrepancies between the global spectrum and the thermodynamic map. For both energy bands, we obtain a best-fitting value of $0.92 \pm 0.02$ on the overall normalization parameter. The C-stat/dof values associated with the 8T fits indicate better fits than the 1T model fits, but not as well as the 2T and 1T $+$ IC model fits (see their C-stat/dof values in Tables \ref{table: model_comparison} and \ref{table: model_comparison_4to20}).

The 8T model represents a good description of the thermal structure of the cluster. Thus, by adding a power-law component to the 8T model, we should, in principle, set a better constraint on the non-thermal emission of the cluster. By adding a power-law component to the 8T model with a fixed/free photon index, we carried out the fits with the data from 3--20 keV and 4--20 keV, as is done in Section \ref{sec: global}. Our findings suggest that the addition of a power-law component improves the fit significantly relative to the 8T model alone (see their C-stat/dof values in Tables \ref{table: model_comparison} and \ref{table: model_comparison_4to20}). However, the 8T $+$ IC model does not reproduce the data as well as the 2T and 1T $+$ IC models. In the 3--20 keV spectral fitting, the best fit yielded a 20--80 keV flux of around 1.5--1.8 $\times 10^{-12}$ erg s$^{-1}$ cm$^{-2}$ on the non-thermal component, whereas the 20--80 keV flux raises to almost double this value in the 4--20 keV spectral fitting.

\section{Discussion}
\label{sec: discussion}
\subsection{Non-thermal component}
\label{sec: IC}
Based on a detailed spectral analysis of deep \textit{NuSTAR} observations of the galaxy cluster SPT J2031, the C-stat/dof values associated with the 3--20 keV spectral fitting (Table \ref{table: model_comparison}) suggest that the 1T $+$ IC model with a free photon index provides a relatively better fit to the global spectrum of the cluster than the models that only include thermal components (1T and 2T). However, the 1T $+$ IC model fit to the 3--20 keV data has an unphysically low-thermal component at 2.1 keV, implying that the SPT J2031 cluster with a mass of around $8 \times 10^{14}$ M$_\odot$ is actually that cold (or not actually that massive). Moreover, the emission measure of the thermal component in the 1T $+$ IC model is lower by a factor of around 1.5 than the emission measure of the IC component, implying that X-ray emission at all energies is dominated by non-thermal emission. Therefore, this model is not physically realistic.

The low-temperature component in the 2T and 1T $+$ IC (free photon index) models might be caused by a slight excess of low energy photons below 4 keV that the low-temperature component may be fitting to (see Fig. \ref{fig: 1T_3to20}). If that's the case, then the values of the best-fitting parameters are expected to be quite different in the 2T and 1T $+$ IC models when the fits are repeated but only with data from 4--20 keV. Indeed, in the 4--20 keV bandpass, we found that the temperature measurements of low-thermal component in the 2T and 1T $+$ IC (free photon index) models increase, respectively, to $5.3_{-0.9}^{+1.6}$ keV and $5.9_{-1.0}^{+2.1}$ keV (Table \ref{table: model_comparison_4to20}), which is around $1\sigma$ lower than the temperature obtained
from the 1T model. These measurements are mostly consistent with the lower-limit temperatures obtained from the 1T model fits to individual regions (Table \ref{table: single_temp_model}).

As shown in Fig. \ref{fig: global_spectrum}, when we allowed the photon spectral index of the non-thermal component to be a free parameter, we found that the thermal component of the 1T $+$ IC model mimics the lower-temperature component of the 2T model, whereas its non-thermal component mimics the higher-temperature component of the 2T model. This implies that X-ray emission at energies below 10 keV is modelled by the thermal component, while the hard emission above 10 keV is mostly modelled by the non-thermal component. This finding differs from what is found for the Bullet cluster \citep{Wik2014}, where the hard emission is mostly modelled by the thermal component, while the IC component appears to be mimicking the lower-temperature component of the 2T model.

Statistically, in the 4--20 keV bandpass, our results show that the 1T $+$ IC model with a free photon index reproduces the data as well as the 2T model, as inferred by their C-stat/dof values (Table \ref{table: model_comparison_4to20}). This suggests that both scenarios are equally likely. This finding is in agreement with the finding of a recent work by \citet{Tumer2022CL0217} that revealed that 2T and 1T $+$ IC models provide fits of equivalent quality to the spectra at the outskirts of the galaxy cluster CL 0217$+$70. Our results also suggest that the 8T $+$ IC model is favoured, statistically, over the 8T model. However, when it comes to the comparison of the 1T $+$ IC and 8T $+$ IC models, the former model provides a better description of the spectral shape of the cluster. 

\begin{figure}
\begin{center}
\includegraphics[width=1.0\columnwidth]{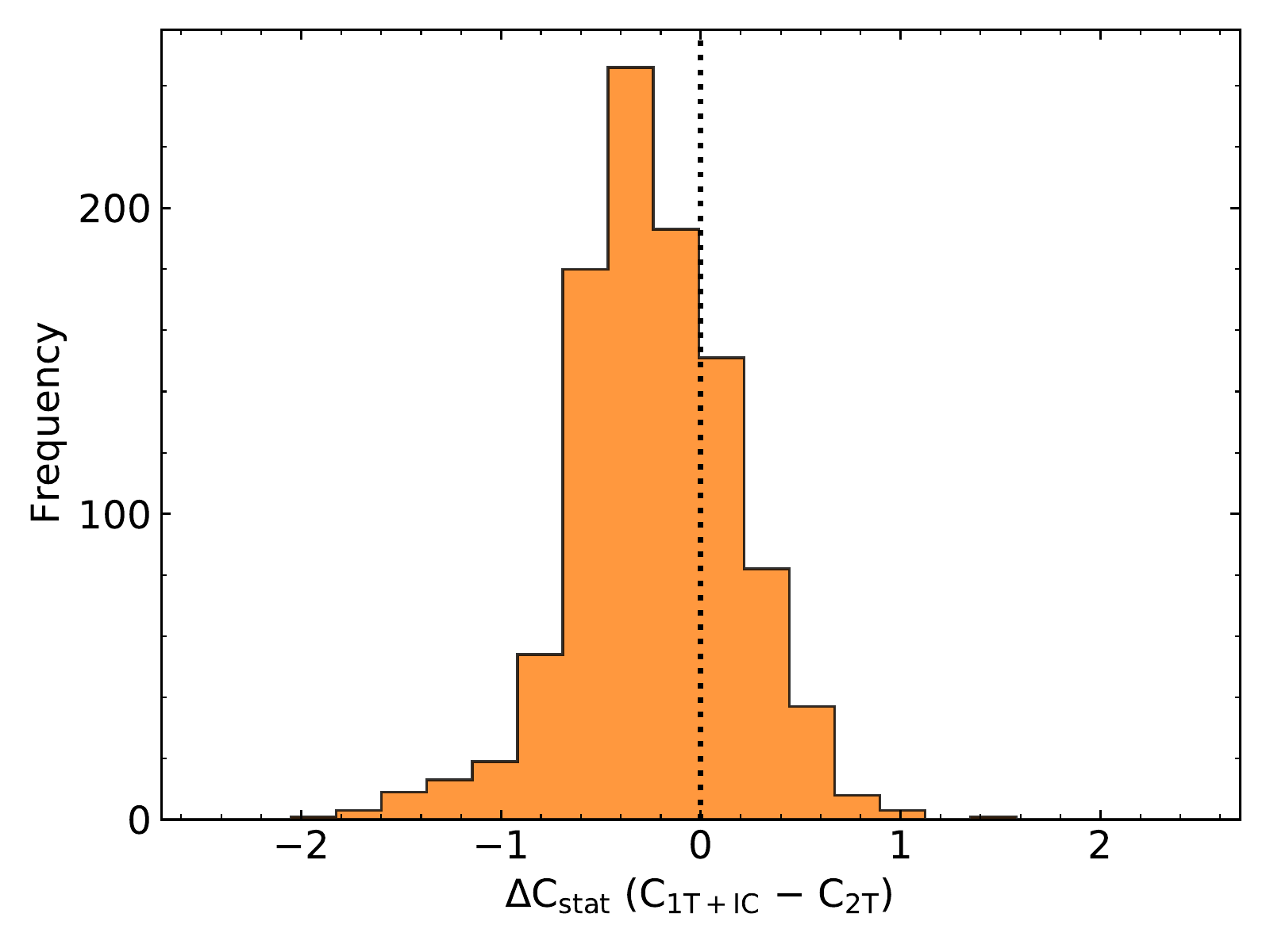}
\end{center}
\caption{Difference of C-stat values between the 1T $+$ IC (free photon index) and 2T models for the 4--20 keV fits using 1000 iterations of the background of each model. The dotted line marks the position where the 1T $+$ IC model with a free photon index and the 2T model provide identical fit statistics to the spectral data. Values to the right of the vertical line show realizations where the 2T model is preferred, whereas values to the left of this line show realizations where the 1T $+$ IC model is preferred. In around 65 per cent of the fits, the 1T $+$ IC model with a free photon index is statistically preferred over the 2T model. }
\label{fig: hist}
\end{figure}



Although the above findings suggest that the 1T $+$ IC model with a free photon index describes the data as well as the 2T model, the relative quality of the 2T versus 1T $+$ IC fits might be affected by the presence of a non-negligible level of systematic fluctuations in the background. To determine which model is preferred with systematic uncertainties due to the background fluctuations included, we calculated the difference in C-stat values between the two best models (2T and 1T $+$ IC with a free photon index) using 1000 realizations of the background that include systematic uncertainties described in Section \ref{sec: spectra}. In Fig. \ref{fig: hist}, we show the difference of C-stat values between the 1T $+$ IC and 2T models from running 1000 iterations of the background of each model. Values to the right of the vertical dotted line show realizations where the 2T model is preferred, whereas values to its left show realizations where the 1T $+$ IC model is preferred. In around 65 per cent of the fits, the 1T $+$ IC model with a free photon index is statistically preferred over the 2T model.

Therefore, our search for IC emission shows a possibility that the hard X-ray emission in the merging galaxy cluster SPT J2031 can be described by a non-thermal component, though a purely thermal origin for the hard emission cannot be completely ruled out. Combining the statistical and systematic uncertainties, the best 1T $+$ IC model fit yields a 20--80 keV flux of $3.93_{-1.10}^{+1.24} \times 10^{-12}$ erg s$^{-1}$ cm$^{-2}$ on the non-thermal IC component. However, the statistical significance of the IC component is marginal (at the $1 \sigma$ level).







\begin{figure}
\begin{center}
\includegraphics[width=1.\columnwidth]{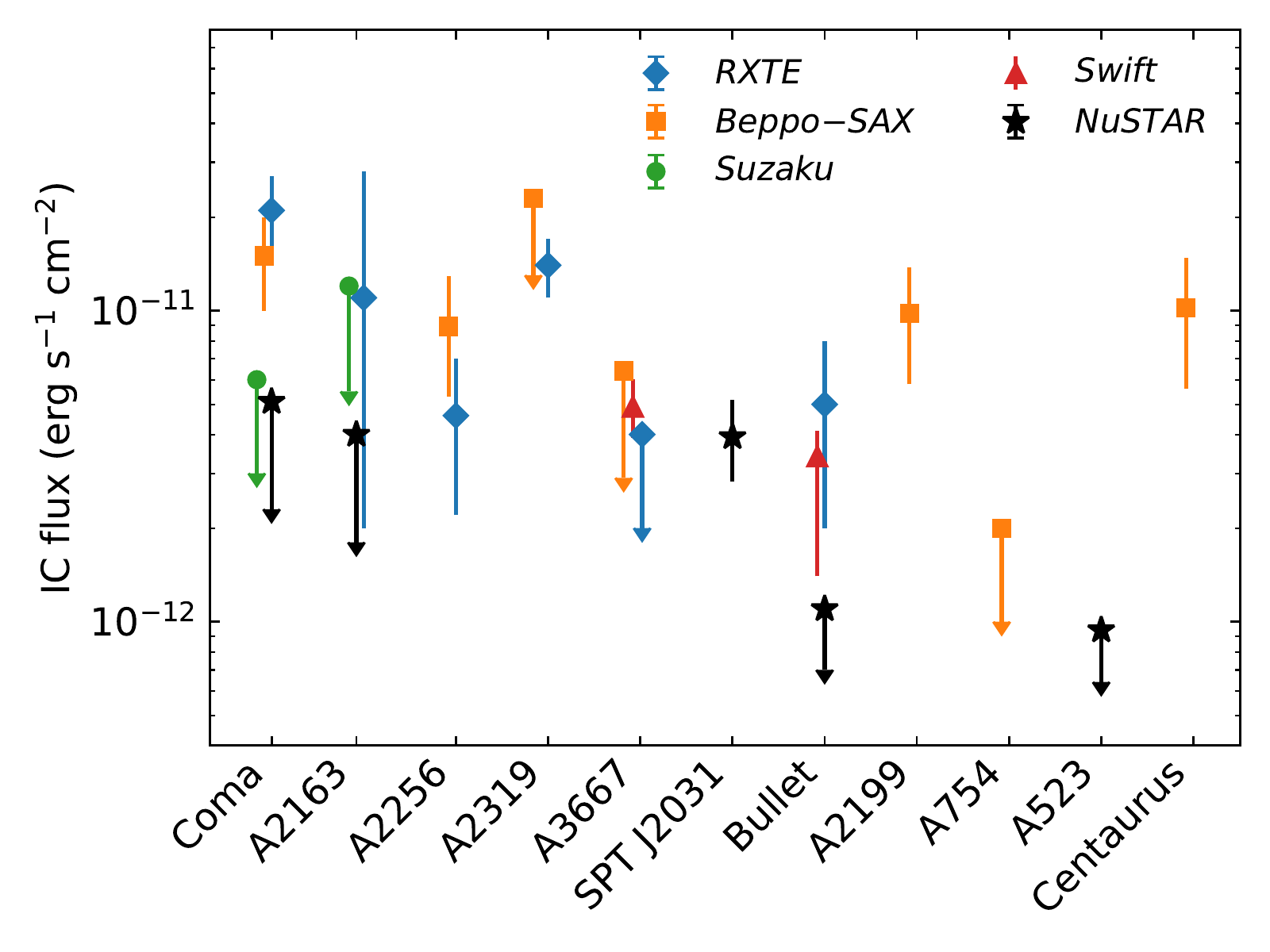}
\end{center}
\vspace{-0.5cm}
\caption{Comparing the non-thermal flux of the SPT J2031 cluster to those reported for other galaxy clusters by various X-ray observatories. }
\label{fig: IC_comparison}
\end{figure}

In Fig. \ref{fig: IC_comparison}, we compare our estimated IC flux for SPT J2031 to those reported for other galaxy clusters using various X-ray observatories. The figure shows that our estimated IC flux is higher than the $90\%$ upper limit IC flux measured by \textit{NuSTAR} for Abell 523 \citep{Cova2019} and the Bullet cluster \citep{Wik2014}, but it is more comparable to the $90\%$ upper limit IC flux estimated for Coma \citep{Gastaldello2015} and Abell 2163 \citep{Rojas2021} using \textit{NuSTAR}. Comparing to the IC flux measurements reported by \textit{RXTE}, \textit{Beppo-SAX}, and \textit{Swift}, the estimated IC flux for SPT J2031 is similar to measurements reported for the Bullet cluster \citep{Petrosian2006,Ajello2010}, Abell 3667 \citep{Fusco-Femiano2001,Rephaeli2004,Ajello2010}, Abell 2256 \citep{Fusco-Femiano2000,Rephaeli2003}, and Abell 2163 \citep{Rephaeli2006}, but it is slightly higher than that estimated for Abell 754 \citep{Fusco-Femiano2003}. Comparing to the IC flux measurements reported by \textit{Suzaku}, our IC flux measurement is consistent with those reported for Coma \citep{Wik2009} and A2163 \citep{Ota2014}. On the other hand, our IC flux falls below that reported for Coma \citep{Rephaeli2002,Fusco-Femiano2004}, Abell 2319 \citep{Molendi1999A2319,Gruber2002}, Abell 2199 \citep{Kaastra1999}, and the Centaurus cluster \citet{Molendi2002} by \textit{RXTE}, and \textit{Beppo-SAX}. The detection significance of non-thermal emission in these galaxy clusters, however, was mostly found to be marginal and controversial \citep[for a review, see][]{Rephaeli2008,Brunetti2014}.


\subsection{Magnetic field}
\label{sec: field}
From the estimated flux of the non-thermal emission, we can directly constrain the average magnetic field strength, assuming that the estimated flux has a non-thermal origin and is produced by the same population of relativistic electrons that emits the radio synchrotron emission. For a power-law energy distribution of electrons, the magnetic field can be estimated as \citep{Govoni2004}: 
\begin{multline}
   B[{\rm{\mu}}G]^{1+\alpha} = h(\alpha) \frac{S_{\rm{R}}[{\rm{Jy}}]}{S_{\rm{X}}[{\rm{erg}}\, {\rm{s}}^{-1} \,{\rm{cm}}^{-2}]} (1+z)^{3+\alpha} \times \\ (0.0545\nu_{\rm{R}}[{\rm{MHz}}])^\alpha \times \big(E_2[{\rm{keV}}]^{1-\alpha} - E_1[{\rm{keV}}]^{1-\alpha}\big),   
    \label{eq: magnetic_field}
\end{multline}
where $h(\alpha)$ is a proportionality constant tabulated in \citet{Govoni2004}, $S_{\rm{R}}$ is the radio flux at frequency $\nu_{\rm{R}}$,  $S_{\rm{X}}$ is the X-ray flux measured over the energy interval $E_1$ and $E_2$, and $\alpha$ is the spectral index.  

Using a total radio flux density of $16.9 \pm 1.8$ mJy at 325 MHz \citep{Raja2020}, we obtain an averaged magnetic field strength of $0.06_{-0.04}^{+0.05}\, \mu \rm{G}$ with the best 1T $+$ IC model fit. When we use a total radio flux density of $232.6 \pm 24.3$ mJy at 150 MHz \citep{intema2017gmrt}, an averaged magnetic field strength of $0.11_{-0.05}^{+0.06} \, \mu \rm{G}$ is obtained. These values are comparable to values of the magnetic field strength found in other galaxy clusters \citep[e.g.][]{Wik2014,Cova2019}. However, our measurements fall below the magnetic field estimated from equipartition condition and  Faraday rotation, which is typically found to be at a level of a few $\mu \rm{G}$ \citep[e.g.][]{Kim1990,Bonafede2010}.

\subsection{Temperature structure}
\label{sec: temp_map}
While the main focus of this work is to determine the character of the hard X-ray emission in the merging galaxy cluster SPT J2031, the temperature map (Fig. \ref{fig: temp_abund_maps}, left-hand panel) exhibits a few significant jumps in the gas temperature, indicating a complex morphology of SPT J2031. There is a very hot region ($\sim 16$ keV with large statistical uncertainties) to the north outside the cluster core. This figure also shows that the gas temperature to the southwest direction drops from $\sim 10$ keV inside the cluster core to $\sim 6$ keV outside the core. The basic properties of these significant jumps in the gas temperature, however, cannot be properly explored using the \textit{NuSTAR} data alone due to \textit{NuSTAR}'s poor spatial resolution. In future work, we will address the nature of these significant jumps in the gas temperature on small scales using deep \textit{Chandra} observations of SPT J2031.

\section{Summary}
\label{sec: summary}
We have carried out a detailed analysis of deep \textit{NuSTAR} observations of the merging galaxy cluster SPT J2031. The main findings of this work can be summarized as follows:
\begin{enumerate}
\item Our findings reveal a possibility that the hard X-ray emission in the SPT J2031 cluster can be described by a non-thermal component, though we cannot completely rule out a purely thermal origin for the hard emission.   
\item With including the statistical and systematic uncertainties, the best model fit yields a 20--80 keV flux of $3.93_{-1.10}^{+1.24} \times 10^{-12}$ erg s$^{-1}$ cm$^{-2}$ on the non-thermal component. This flux is mostly comparable to those found in other galaxy clusters using \textit{NuSTAR}, as well as other X-ray observatories.
\item By combining the estimated IC flux with the synchrotron emission at radio frequencies, we obtain an averaged magnetic field strength in the range of around 0.1--0.2 $\mu$G, averaged over the entire radio-emitting region. This value roughly agrees with those found in other clusters using \textit{NuSTAR}.

While our results reveal a possible non-thermal origin of the hard X-ray emission in the merging galaxy cluster SPT J2031, alternative origins for this hard emission, such as clumps of very hot gas and slightly underestimated backgrounds, are also plausible. However, given its unprecedented focusing capability in the hard X-ray energy band, deep \textit{NuSTAR} observations of additional hot galaxy clusters, such as the one studied here, would be able to set a more robust constraint on the high-energy emission in galaxy clusters and to address some of the unanswered questions related to this emission.

\end{enumerate}

\section*{Acknowledgements}
We thank the referee, Daniel Wik, for the very helpful and informative report. We acknowledge support from the NASA \textit{NuSTAR} grant 80NSSC21K0074. This research made use of data from the \textit{NuSTAR} mission, a project led by the California Institute of Technology, managed by the Jet Propulsion Laboratory, and funded by NASA. We thank the \textit{NuSTAR} Operations, Software, and Calibration teams for support with the execution and analysis of these observations. This research has made use of the NuSTAR Data Analysis Software \textsc{Nustardas} jointly developed by the ASI Science Data Center (ASDC, Italy) and the California Institute of Technology (USA).

\section*{Data Availability}
This work is based on the \textit{NuSTAR} data publicly available through the \textsc{heasarc} Archive (https://heasarc.gsfc.nasa.gov). The software packages \textsc{HEASoft} and \textsc{xspec} were used, and these can be downloaded from the \textsc{heasarc} software web page. Analysis and figures were produced using \textsc{python} version 3.8.



\bibliographystyle{mnras}
\bibliography{spt_cl2031} 







\bsp	
\label{lastpage}
\end{document}